\documentclass[page-classic]{epl2}
\usepackage{epsfig}
\usepackage{soul}
\usepackage[normalem]{ulem}
\usepackage[autostyle]{csquotes}
\usepackage{amsmath}

\title{Dynamics of cholesteric liquid crystals in the presence of 
random magnetic fields}

\shorttitle{Stochastic dynamics of cholesteric liquid crystal}

\author{Amit K Chattopadhyay\inst{1} \thanks{E-mail 
address: a.k.chattopadhyay@aston.ac.uk} \and Prabir K. Mukherjee\inst{2}} 
\shortauthor{A. K. Chattopadhyay and P. K. Mukherjee}

\institute{
\inst{1}Department of Mathematics, Aston University,
Birmingham B4 7ET, England \\
\inst{2}Department of Physics, Government College of Engineering
and Textile Technology, 12 William Carey Road, Serampore, Hooghly-712201,
India}

\abstract{Based on dynamic renormalization group techniques, this letter analyzes the effects of external stochastic perturbations on the dynamical properties of cholesteric liquid crystals, studied in presence of a random magnetic field. Our analysis quantifies the nature of the temperature dependence of the dynamics; the results also highlight a hitherto unexplored regime in cholesteric liquid crystal dynamics. We show that stochastic fluctuations drive the system to a second-ordered Kosterlitz-Thouless phase transition point, eventually leading to a Kardar-Parisi-Zhang (KPZ) universality class. The results go beyond quasi-first order mean-field theories, and provides the first theoretical understanding of a KPZ phase in distorted nematic liquid crystal dynamics.
}


\begin{document}

\maketitle

\newpage

A liquid crystal is a special state of matter that shares properties both from a liquid as well as from a (solid) crystal. They are differentiated based on their optical properties, that, in turn, depend both on positional ordering as well as on orientational ordering. 
In effect, a liquid crystal sample exhibits a competition between the alignment of the director ${\bf n}$ at a surface or boundary and the orientation induced within the sample by an externally applied electric or magnetic field which produces a spatial rearrangement of the director subjective of the phase concerned \cite{degennes,priestly,chandrasekhar}. 
A cholesteric (Ch) mesophase is locally very similar to a nematic phase
except that it is composed of optically active molecules. When an optically active
material forms a nematic phase, the preferred direction of the long-molecular
axis varies in a direction throughout the medium in a regular way and displays
a continuous twist along the optical axis leading to a helical structure. Thus
the cholesteric phase is a distorted nematic phase. The present article provides the first theoretical model
addressing the phase transition properties of a cholesteric liquid crystal perturbed
by a stochastic magnetic field.

As noted before, a liquid crystal is characterized by the configurational deformation driven by a magnetic field. In thermodynamic parlance, this is defined as a Freedericksz transition that occurs in a liquid crystal
slab when the director reorientates in the direction of an applied 
magnetic field larger than a critical field $H_C$. It is also well known that the 
Freedericksz instabilities occur in nematic liquid crystals 
\cite{degennes,priestly,chandrasekhar,stewart,stephen,brochard}. Planar cholesteric films can also exhibit
either a Freedericksz transition or a transition to a stripe texture in 
electric fields \cite{chigrinov}. Similar Freedericksz instabilities are expected in cholesteric liquid crystals.
Valkov et al \cite{valkov} studied the 
Freedericksz transition in the twist cells of cholesteric liquid crystals with a 
finite surface energy. It was shown that this transition can be 
either of the second order or of the first order depending on the values of 
the Frank constants, pitch, surface energies, and the cell thickness 
\cite{valkov}. Zumer et al \cite{zumer} theoretically analyzed 
the field induced structural (Freedericksz) transitions in supramicron nematic 
and cholesteric liquid crystal droplets embedded in a solid matrix. They 
obtained the equilibrium structures and phase diagrams by taking into account 
the elastic, surface, and field free-energy contributions. 
Experimental work
on dynamics of Freedericksz transition \cite{lonberg,hui,guyon} shows the
existence of transient spatial structures.
Sagues and Miguel \cite{sagues1} studied the Freedericksz transition under 
twist deformation in a nematic layer when the magnetic field has 
a random component. It was found that the randomness of the field produces a 
shift in the instability point. Beyond this instability point, the time 
constant characteristic of the approach to the stationary stable state 
decreases because of the field fluctuations.  
 
In a remarkable departure from deterministic analysis, Miguel \cite{miguel1} studied the stationary spatial correlations of the
director field of a nematic liquid crystal near the Freedericksz transition in
a magnetic field with a time-dependent random component. 
They showed that the correlation length diverges at a shifted instability point.
Sagues and Miguel \cite{sagues2} studied transient dynamics of spatial 
fluctuations of the director field in the pure twist Freedericksz transition. 
They analyzed linear 
theories at different stages of evolution and the domain of validity.
In a later work \cite{miguel2}, the same authors studied the dynamics of a transient pattern 
formation in the Freedericksz transition corresponding to a twist geometry.  
They calculated the time-dependent structure factor based on a dynamical model 
which incorporates consistently the coupling of the director field with the 
velocity flow and also the effect of fluctuations. They identified a  
time scale associated with the emergence of the pattern and a 
a subsequent stage of pattern development.

Apart from rare theoretical attempts routed through renormalization group applications \cite{amit}, none of these aforementioned studies, however, analyze dynamic perturbations 
of cholesteric liquid crystals in presence of random magnetic field, as is often seen or expected in 
experiments \cite{lonberg,hui,guyon}.
The purpose of the present paper is to discuss
the dynamical evolution patterns of cholesteric liquid crystals in a random magnetic
field using the language of dynamic renormalization groups.
We elucidate the main
consequences of a fluctuating environment in such dynamical aspects.
We show that in the experimentally viable low-frequency limit, that is at large spatiotemporal scales, the
dynamical model predicts the onset of a Kosterlitz-Thouless transition from the emergence of an equivalent sine-Gordon
model, in agreement with results mapped from different phenomenological models \cite{akc,rost}. At large enough spatiotemporal scales, the inherent nonlinearity drives the system to a 
Kardar-Parisi-Zhang fixed point, represented by appropriate self-affine scaling.
  
We start with a Frank free energy \cite{frank}, consistent with the phenomenological continuum theory of liquid 
crystals that is very successful in explaining various magnetic or electric field-induced
effects. When the external field is applied perpendicularly to the helical 
axis, at low temperatures, it unwinds the helical structure of the cholesteric
state via a nucleation-type transition \cite{degennes}.
The elastic free energy of a cholesteric liquid crystal in the 
presence of magnetic field can be written as \cite{stephen}

\begin{equation}
g_{ch}=\frac 12K_1({\bf \nabla}\cdot{\bf n})^2+\frac 12K_2({\bf n}\cdot({\bf \nabla}
\times{\bf n})+q_0)^2+\frac 12 K_3({\bf n}\times({\bf \nabla}\times{\bf n}))^2
-\frac 12\chi_a ({\bf n}\cdot {\bf H})^2
\label{free1}
\end{equation}
where $K_{i}$ (i=1,2,3) are three elastic constants associated with splay, twist and
bend deformations respectively. $\chi_a$ is the anisotropic part of the 
magnetic susceptibility and $q_0$ is the wave vector corresponding
to the magnetic field amplitude $H=0$ ($q_0=2\pi/\lambda_0)$, $\lambda_0$ being the pitch of the helix. In the rest of this article, we will always consider positive anisotropy magnetic susceptibility.

Let us start with a cholesteric sample
of thickness $d=\lambda_0/2$.
We assume that, at the walls strong anchoring prevails along the x-axis, which then becomes the direction of the applied external magnetic field ${\bf H}$. The helical axis z is chosen
normal to ${\bf H}$. This defines pure
twist with the following components of the director $\bf n$ orientation:
$n_x=\cos\phi(z,t)$, $n_y=\sin\phi(z,t)$. $n_z=0$. $\phi(z,t)$ is the angle
between $\bf n$ and x axis. Boundary conditions are $\phi(z=0)=0$ and
$\phi(z=d)=\pi$. Then the non-holonomic elastic free energy $G$ is given by

\begin{equation}
G=\frac S2 \int^d_0\left[\frac {K_2}{2}\left(\frac {d \phi}{dz}-q_0\right)^2
-\frac {\chi_aH^2}{2}\cos^2\phi\right]\:dz = \int L(z,\phi(z),\phi^{\prime}(z))\:dz,
\label{free2}
\end{equation}
where $S$ is the surface area of the plates and $L(z,\phi(z),\phi^{\prime}(z))$ ($\phi^\prime(z)=\frac{d\phi}{dz}$) is the equivalent Lagrangian of the system.

The time dependent Ginzburg-Landau (TDGL)-type model can be specified by the following Langevin dynamics

\begin{equation}
M\frac {\partial \phi(z,t)}{\partial t}=-\frac {\delta G}{\delta \phi} + F_0
+\eta(z,t),
\label{lang1}
\end{equation}

where $\eta(z,t)$ is a stochastic white noise represented by $<\eta(z,t)\: \eta(z',t')> = D_0 \:\delta(z-z') \delta(t-t')$ and $<\eta(z,t)>=0$, in which the curly brackets \enquote{$< >$} represent ensemble average over all noise realizations and $D_0$ is the noise strength, a constant, while $\frac{\delta G}{\delta \phi}$ is the functional derivative of $G$ with respect to the variable $\phi$. $F_0$ is a constant \enquote{overdamped force} that pumps a steady energy into the system. While admittedly the original Frank-Oseen model (Frank energy) does not have such a constant force ($F_0$) in its definition, such a term is related to the existence of a finite non-trivial kinetic roughening in the model that can only be counterbalanced by a negative damping term $-F_0$. This will be later analyzed in detail through a renormalization group analysis. 

Equation (\ref{lang1}) can then be rewritten as
\begin{equation}
M\frac {\partial \phi}{\partial t}=\frac{S}{2}\left(-\frac {\chi_aH^2}{2}\sin2\phi+K_2\nabla^2
\phi \right)+F_0+\eta(z,t)
\label{lang3}
\end{equation}
The above equation (\ref{lang3}) represents a quintessential sine-Gordon model \cite{nozieres} whose low frequency spatiotemporal scaling properties have been studied by Chattopadhyay\cite{akc}. In what follows, we will consistently use $\nabla^2 \phi=\dfrac{\partial^2 \phi}{\partial z^2}$ in order to restrict our analysis to the axial function.

Along the $z$-direction, the steady-state dynamics of equation (\ref{lang3}) leads to
\begin{equation}
K_2\frac {d^2\phi}{dz^2}=\frac {\chi_aH^2}{2}\sin2\phi - \frac{2F_0}{S}
\label{deter}
\end{equation}
which, for $F_0=0$, can be easily solved in terms of Jacobi elliptic functions. For a general non-zero $F_0$, equation (\ref{deter}) may be numerically solved.

Equation (\ref{lang3}) can be further rescaled, which, on a periodic lattice, leads to the following model

\begin{equation}
\eta\frac {\partial h}{\partial t}=\gamma \frac {d^2h}{dz^2}-\frac {2\pi V}
{a}\sin\left(\frac {2\pi}{a}h\right)+F+N(z,t)
\label{lang5}
\end{equation}
where $\eta=\frac {M\pi}{aK_2}$, $\frac {2\pi V}{a}=\frac{S}{4\xi^2}$, $\gamma=\frac{\pi S}{2a}$, $F=\frac{F_0}{K_2}$, $\xi^2=\frac{K_2}{\chi_a H^2}$ and the rescaled noise $N(z,t)=\eta(z,t)/K_2$. The above is the classic noisy sine-Gordon  
model that can be studied using the method previously used by
Chattopadhyay \cite{akc}. 

The model presented in Equation (\ref{lang5}) is the d=1+1 dimensional
Rost-Spohn model \cite{rost} or Nozieres-Gallet model \cite{nozieres} for
$\lambda=0$. 
The second moment of the noise-noise correlation in the wave-vector space is then given by
\begin{equation}
<N(k,t)N(k^{\prime},t^{\prime})>=2Du\left(\frac {\pi k}{\Lambda}\right)\delta (k+k^{\prime})\delta(t-t^{\prime})
\label{noise}
\end{equation}
where $u(1-x)=\theta(x)$, a Heaviside step function, $\Lambda\propto\frac {\pi}{a}$ and $D=\eta T$. 
$a$ is taken to be a lattice constant in abeyance with normalization within a rectangular box of sides $a$ each, essentially to map this to the Chattopadhyay-model presented in \cite{akc}.

In the following analysis, we employ standard dynamic renormalization group 
(RG) techniques to analyze the model dynamics in line with the Chattopadhyay \cite{akc} approach. The fast frequency components are integrated over the momentum shell $\Lambda e^{-dl}<|k|<\Lambda$ 
i.e. $\Lambda (1-dl)$ for $<|k|<\Lambda$, in the limit of infinitesimally small $dl$. We assume $V$ to be a perturbative 
constant. Now the effective Frank free energy 
functional for zero driving force ($F=0$ and $N=0$) can be written as 

\begin{equation}
G=\int dz \Big[ \frac {\gamma}{2}(\nabla h-q_0)^2-V\cos(\frac {2\pi}{a}h) \Big],
\label{free3}
\end{equation}

where we rescaled the renormalization variables $k\rightarrow k^{\prime}=
(1+dl)k$, $z\rightarrow z^{\prime}=(1-dl)z$, $h\rightarrow h^{\prime}=h$,
$t\rightarrow t^{\prime}=(1-2dl)t$, $\eta\rightarrow \eta^{\prime}=\eta$,
$\gamma\rightarrow \gamma^{\prime}=\gamma$, 
$V\rightarrow V^{\prime}=(1+2dl)V$ and $F\rightarrow F^{\prime}=(1+2dl)F$

First we discuss the perturbative dynamics. In this case equation (\ref{lang5}) 
can be rewritten as
\begin{equation}
\eta\frac {\partial h}{\partial t}=\gamma \nabla^2 h+\Psi(h)+N
\label{lang6}
\end{equation}
where $\Psi(h)=-\frac {2\pi V}{a}\sin[\frac {2\pi}{a}(h+\frac{Ft}{\eta})]$.

Using perturbation expansions for the dynamic variables $X_i=h, N$, we can write $X_i=\bar{X_i}+\delta X_i$, where the quantity $\bar{X}$ is defined within the momentum range $|k|<(1-dl)\Lambda$, with $\delta X$ defined inside the annular ring 
$(1-dl)\Lambda<|k|<\Lambda$, we get

\begin{subequations}
\begin{equation} 
\eta\frac {\partial \bar{h}}{\partial t}=\gamma \nabla^2 \bar{h}+\bar{\Psi}(\bar{h},\delta h)+\bar{N},\:\:
\label{lang7}
\end{equation}
\begin{equation}
\eta\frac {\partial \delta h}{\partial t}=\gamma \nabla^2 \delta h+\delta \Psi(\bar{h},\delta h)+\delta N,
\label{lang7}
\end{equation}
\end{subequations}
where $\bar{\Psi}=<\Psi>_{\delta N}$, averaging defined over all noise perturbations.
$\bar{\Psi}$ and $\delta {\Psi}$ can be expressed as 

\begin{equation}
\bar{\Psi}=-\frac {2\pi V}{a}\sin[\frac {2\pi}{a}(\bar{h}+\frac{F t}{\eta})]
(1-\frac {2\pi^2}{a^2}<\delta h^2>_{\delta N}),\:\:
\delta {\Psi}=-\frac {4\pi^2 V}{a^2}\cos[\frac {2\pi}{a}(\bar{h}+\frac{F t}{\eta})] \delta h.
\label{npsi1}
\end{equation}

Expanding $\delta h=\delta h^{(0)}+\delta h^{(1)}+....$ perturbatively, we get

\begin{subequations}
\begin{equation}
\delta h^{(0)}(z,t)=\int^t_{-\infty}dt^{\prime}\int dz^{\prime}\: G_0(z-z^{\prime},
t-t^{\prime})\:\delta N(z^{\prime},t^{\prime}),
\label{expand1}
\end{equation}
\begin{equation}
\delta h^{(1)}(z,t)=\int^t_{-\infty}dt^{\prime}\int dz^{\prime}\: G_0(z-z^{\prime},
t-t^{\prime})\times[-\frac {4\pi^2 V}{a^2}\cos[\frac {2\pi}{a}(\bar{h}
(z^{\prime},t^{\prime})+\frac{Ft^{\prime}}{\eta})]\: \delta h^{(0)},
\label{expand2}
\end{equation}
\end{subequations}

where $G_0(x,t)=\frac {1}{2\pi \gamma t}e^{-\frac {\eta x^2}{2\gamma t}}$ is 
the Green's function. For the initial condition $h(z,t=-\infty)=0$, we get 

\begin{subequations}
\begin{equation}
<{(\delta h)}^2(z,t)>=<{(\delta h^{(0)})}^2>+2<\delta h^{(0)}(z,t)\:
\delta h^{(1)}(z,t)>,
\label{expand3}
\end{equation}
\text{and}
\begin{equation}
<(\nabla \delta h)^2(z,t)>=<(\nabla \delta h^{(0)})^2>+2<\nabla \delta 
h^{(0)}\:
\nabla \delta h^{(1)}>.
\label{expand4}
\end{equation}
\end{subequations}

From equations (\ref{expand1})-(\ref{expand4}), we can now write
\begin{equation}
\delta h_k(t)=\frac {1}{\eta}e^{-\frac {\gamma}{\eta} k^2 t}\int_0^tdt^{\prime}
\delta N_k(t^{\prime})e^{\frac {\gamma}{\eta} k^2 t^{\prime}}
\label{lang9}
\end{equation}
The corresponding correlation functions are given by

\begin{subequations}
\begin{equation}
<\delta h^{(0)}(z,t)\:\delta h^{(0)}(z^{\prime},t^{\prime})>=\frac {\eta T}
{2\pi \gamma}e^{i\Lambda |z-z^{\prime}|}\:e^{-\frac {\gamma}{\eta}\Lambda^2
|t-t^{\prime}|} dl
\label{corre3}
\end{equation}
\begin{equation}
<\partial_i\delta h^{(0)}(z,t)\:\partial_j \delta h^{(0)}(z^{\prime},t^{\prime})>=\frac {\eta T}
{2\pi \gamma}e^{i\Lambda |z-z^{\prime}|}\:e^{-\frac {\gamma}{\eta}\Lambda^2
|t-t^{\prime}|}\:(t\Lambda^2)\:dl\:\delta_{ij}
\label{corre4}
\end{equation}
\end{subequations}

We now calculate the RG flow equations correct up to the second order

\begin{subequations}
\begin{equation}
dV^{(1)}=-\frac {2\pi^2}{a^2}<\delta h^{(0)2}(z,t)>=\frac {\pi\eta T}{\gamma a^2}dl,
\label{first1}
\end{equation}
\begin{equation}
dF^{(1)}=0
\label{first2}
\end{equation}
\end{subequations}

Then it is easy to calculate the first order renormalization group (RG) flows for $V$ and $F$. In 
order to calculate the RG flows for $\gamma$, $\eta$ and $D$, we need to 
evaluate the second ordered corrections terms $<\delta h^{(0)}\delta h^{(1)}>$ and 
$<\partial_i\delta h^{(0)}\partial_j\delta h^{(1)}>$. These two terms can be 
expressed as

\begin{subequations}
\begin{eqnarray}
<\delta h^{(0)}(z,t)\delta h^{(1)(z,t)}>&=&-\frac {4\pi^2}{a^2}V\int_{-\infty}^tdt^{\prime}
\int dz^{\prime} \cos[\frac {2\pi}{a}(\bar{h}+\frac {Ft^{\prime}}{\eta})]
G_0(z-z^{\prime},t-t^{\prime}) \nonumber \\
&&\times<\delta h^{(0)}(z,t)\delta h^{(0)}(z^{\prime},t^{\prime})>
\label{ncorre1}
\end{eqnarray}
\text{and}
\begin{eqnarray}
<\partial_i \delta h^{(0)}(z,t)\partial_i \delta h^{(1)}(z,t)>&=&-\frac {4\pi^2}{a^2}V
\frac {\eta}{2\gamma}\sum_{i=1}^2\int_{-\infty}^tdt^{\prime}
\int dz^{\prime} \cos[\frac {2\pi}{a}(\bar{h}(z^{\prime},t^{\prime})+\frac {Ft^{\prime}}{\eta})]\nonumber \\
&&\times\left(\frac {z_i^{\prime}-z_i}{t-t^{\prime}}\right) 
G_0(z-z^{\prime},t-t^{\prime}) \\ \nonumber
& \times & <\partial_i \delta h^{(0)}(z,t)\:\partial_i \delta h^{(0)}(
z^{\prime},t^{\prime})>.
\label{ncorre2}
\end{eqnarray}
\end{subequations}

Combining the variable rescaling $t\rightarrow t^{\prime}=t/\eta$ and
$h\rightarrow h^{\prime}=h-\frac {Ft}{\eta}$ with the Rost-Spohn
approximation scheme, the second-ordered renormalized
$<\delta h^{(0)}(z,t)\:\delta h^{(1)}(z,t)>$ gives

\begin{eqnarray}
\Psi_{SG}&\approx &\displaystyle -\frac {8\pi^3V^2T}{\gamma^2a^5}dl\int_{-\infty}^t\frac {dt^{\prime}}
{t-t^{\prime}}\int dz^{\prime}e^{i\Lambda|z-z^{\prime}|} \times e^{-[\frac {\gamma}{2\eta}\frac {(z-z^{\prime})^2}{(t-t^{\prime})}-\frac{\gamma}
{\eta}\Lambda^2(t-t^{\prime})
-\frac {2\pi^2}{a^2}{<(\bar{h}(z,t)-\bar{h}
(z^{\prime},t^{\prime})^2>}_N]} \nonumber \\
&\times &\bigg[\frac {2\pi}{a}\left(\frac {\partial }{\partial t}\bar{h}(z,t)(t-t^{\prime})-
\frac 12 \partial_i\partial_j\bar{h}(z,t)(z_i-z_i^{\prime})(z_j-z_j^{\prime})\right)
 \nonumber \\
&\times &\cos\left(\frac {2\pi}{a}\frac {F}{\eta}(t-t^{\prime})\right)+\left(1-\frac {2\pi^2}{a^2}[
\partial_i \bar{h}(z,t)]^2(z_i-z_i^{\prime})^2\right) \times \sin\left(\frac {2\pi}{a}\frac {F}{\eta}(t-t^{\prime})\right) \bigg].
\label{npsi5}
\end{eqnarray}
The terms in equation (\ref{npsi5}) that are respectively proportional to 
$\frac {\partial \bar{h}}{\partial t}$, $\partial_i\partial_j\bar{h}$ and 
$(\partial_i\bar{h})^2$ are the renormalized contributions for $\eta$, $\gamma$ 
and a new KPZ nonlinearly $\lambda$ that automatically gets created from a
sine-Gordon potential. The constant term above renormalizes $F$. 

This leads to the RG flow equations as follows:

\begin{subequations}
\begin{equation}
\frac {dU}{dl}=(2-n)U,
\label{flow3}
\end{equation}
\begin{equation}
\frac {d\gamma}{dl}=\frac {8\pi^4}{\gamma a^4}n A^{(\gamma)}(n;k)U^2,
\label{flow4}
\end{equation}
\begin{equation}
\frac {d\eta}{dl}=\frac{32\pi^4}{\gamma a^4}\dfrac{\eta}{\gamma} nA^{(\eta)}
(n;\kappa)U^2
\label{flow5}
\end{equation}
\begin{equation}
\frac {d\lambda}{dl}=\frac {32\pi^5}{\gamma a^5}nA^{(\lambda)}(n,\kappa)U^2,
\label{flow6}
\end{equation}
\begin{equation}
\frac {dD}{dl}=\frac{32\pi^4}{\gamma a^4}\dfrac{D}{\gamma} nA^{(\eta)}
(n;\kappa)U^2
\label{flow7}
\end{equation}
\begin{equation}
\frac {d K}{dl}=2K-\frac {8\pi^3}{\gamma a^3}nA^{(K)}(n;\kappa)U^2,
\label{flow8}
\end{equation}
\end{subequations}

where $U=V/\Lambda^2$, $K=F/\Lambda^2$, $\bar{\rho}=\Lambda \rho$, $x= \dfrac {\gamma(t-t^{\prime})}{\eta \rho^2}$, 
$n=\dfrac{\pi \eta T}{\gamma a^2}$ and $\kappa=\dfrac {2\pi K}{a\gamma}$.

The functional forms of the renormalization group (RG) flow integrals $A^{(\gamma)}(n;\kappa)$, $A^{(\eta)}(n;\kappa)$, $A^{(\lambda)}(n;\kappa)$ and $A^{(K)}(n;\kappa)$ are defined as follows:

$A^{(\gamma)}(n;\kappa)=\displaystyle \int_0^{\infty}\frac {dx}{x}\int_0^{\infty}d\bar{\rho}\:
\bar{\rho}^2\:\cos({\bar{\rho}})\:\cos\left(\frac {2\pi}{a}\frac {Kx\bar{\rho}^2}{\gamma}\right)\:e^{-f}$,

$A^{(\eta)}(n;\kappa)=\dfrac {x}{\bar{\rho}^2}A^{(\gamma)}(n;\kappa)$,

$A^{(\lambda)}(n;\kappa)=\displaystyle \int_0^{\infty}\frac {dx}{x}\int_0^{\infty}d\bar{\rho}\:
\bar{\rho}^2\:\cos({\bar{\rho}})\:\sin\left(\frac {2\pi}{a}\frac {Kx\bar{\rho}^2}{\gamma}\right)\:e^{-f}$,

$A^{(K)}(n;\kappa)=\dfrac{1}{\bar{\rho}^2}A^{(\lambda)}(n;\kappa)$,

$f=\frac {1}{2x}+x\bar{\rho}^2+\dfrac {2\pi \eta}{\gamma a^2}\:\chi(\bar{\rho},x)$,

$\chi(|z-z'|,|t-t'|)=\dfrac {\pi\gamma}{\eta T} <[\bar{h}(z,t)-\bar{h}(z^{\prime},t^{\prime})]^2>_{\bar{N}}$,

$<[\bar{h}(z,t)-\bar{h}(z^{\prime},t^{\prime})]^2>_{\bar{N}}=\dfrac{\eta T}{\pi \gamma}\displaystyle \int_0^1dk\big[\sqrt{2(1-\cos{k|z-z'|})}\big]\:e^{-\frac{\gamma}{\eta}k^2(t-t')}$.

\begin{center}
\begin{figure}[tbp]
\includegraphics[height=5.5in,width=3in,angle=-90]{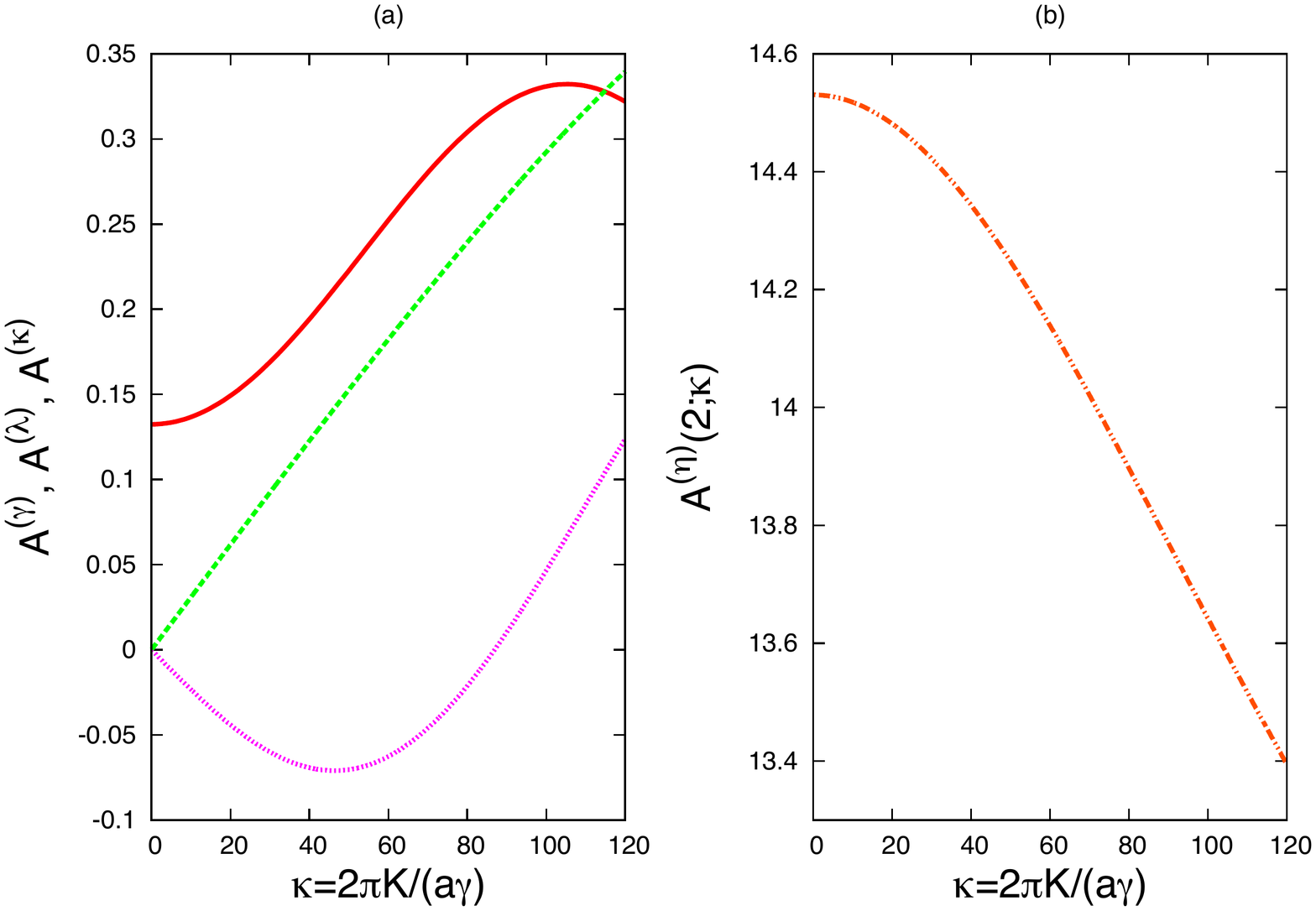}
\caption{Variation of the RG flow integrals. Plot (a) shows the variations of $A^{(\gamma)}(2;\kappa)$, $A^{(\lambda)}(2;\kappa)$ and $A^{(\kappa)}(2;\kappa)$ against $\kappa$ while plot (b) shows the variation of $A^{(\eta)}(2;\kappa)$ with $\kappa$.
\label{fig_Aglk}}
\end{figure}
\end{center}

We now discuss the physical realization of the RG flow equations. 
The variation of the flow patterns at the fixed point $n=\dfrac{\pi \eta T}{\gamma a^2}=2$ are depicted in Figure \ref{fig_Aglk}; this clearly suggests that as the external force $K$ (or proportionately $F$) increases, both the renormalized diffusion constant $\gamma$ and the forcing strength increase. Since $\gamma \sim \frac{1}{\sqrt{\eta T}}$, as $\gamma$ increases, the non-equilibrium temperature of the system decreases and vice versa. The most interesting aspect of these flows, though, is the emergence of a hitherto unseen Kardar-Parisi-Zhang (KPZ) \cite{kpz} nonlinearity that steadily increases with the dynamical flow thereby leading to an eventual KPZ fixed point. 


In complementarity, the friction term $\eta$ gets compressed in the flow space resulting in a decay curve as shown in Figure \ref{fig_Aglk}(b). Although we do not have any $({\bf \nabla} \phi)^2$ term (KPZ
term) in the original dynamics as shown in equation (\ref{lang3}), such a term is generated due to the presence of the sine-Gordon potential in the dynamics. Even if
$\lambda=0$ at the start, the $\frac {d\lambda}{dl}$ in equation (\ref{flow6})
ensures that a nonzero $\lambda$ is eventually generated. Once $\lambda \ne 0$,
effective temperature of the system increases due to renormalization.
When $n=\frac {\pi \eta T}{\gamma a^2}>2$, $U$ decays to zero. Thus at a large
enough spatiotemporal scale, the dynamics is governed by KPZ scaling.

The prediction of the KPZ universality class for 1+1 dimensions validates a recent experimental work by 
Takeuchi and Sano \cite{sano} where they studied the scale-invariant 
fluctuations of growing interfaces in nematic liquid crystal turbulence.
As a perturbed nematic phase, a cholesteric phase too is expected to behave identically.

The flow equations given by equations (\ref{flow3}-\ref{flow8}) are not valid at
very small $T$'s.
Although the original model (Frank energy) does not have a constant force ($F$)
term, such a term will be generated by a renormalization group analysis of the model. In order to control kinetic roughening,
of the cholesteric liquid crystal, especially in the high temperature regime,
an opposing force $-F$ is ingrained into the dynamics at all times. The results from this analysis, thus identify how best to control the dynamics of a cholesteric liquid crystal and what particular combination of parameters is capable of achieving this. Also the results go beyond the usual confines of mean-field analysis, now including the second order RG-correction in analyzing the contribution from the dynamical modes.

In conclusion, we have studied the impact of stochastic fluctuations in modifying the spatiotemporal properties of cholesteric liquid crystals in presence of a random magnetic field. In the mode considered, we focus on strong anchoring of the liquid crystal molecules to the boundaries. For general studies on surface anchoring, a Rapini-Papoular potential may be considered in line with \cite{belyakov}. The core renormalizability aspect of the theory will still remain intact though and would be perfectly applicable with such potentials too. Our results show that stochastic fluctuations 
drive the system to a Kosterlitz-Thouless transition point, through a second 
ordered phase transition, thereby converging to a Kardar-Parisi-Zhang universality class. The nature of the temperature dependence of the cholesteric phase has also been explained. The model additionally provides the first theoretical confirmation of the recently observed KPZ-phase in nematic liquid crystals \cite{sano}.
Our results are expected to encourage further experiments, including the study of the dynamical properties of other liquid crystalline phases, focusing now on finite sized effects of samples, an aspect that we are presently working on.

\end{document}